\documentclass[journal]{IEEEtran}
\usepackage{cite}
\usepackage[pdftex]{graphicx}
\usepackage[caption=false,font=footnotesize]{subfig}
\usepackage{stfloats}
\fnbelowfloat
\usepackage{url}

\usepackage{multirow}
\begin{document}
%
\title{Simulating the Difference between a DES and a Simple Railgun
using SPICE}

\author{
\IEEEauthorblockN{S.\,Hundertmark}\\
\IEEEauthorblockA{French-German Research Institute
of Saint-Louis,
France}
}
\maketitle
\begin{abstract}
A DES railgun increases the launch efficency when compared to a simple
railgun setup. In a {\sc Ngspice}-Simulation a DES and a simple railgun setup 
connected to an existing 10-MJ capacitor based
power supply are compared. Important parameters like velocity, power
and efficiency are used to quantify the performance gain, when using
the mechanically more complex DES setup. The results of this study
were used to select the simple railgun setup as the design for a new barrel
to the 10-MJ power supply.
\end{abstract}

\section{Introduction}
The French-German Research Institute of Saint-Louis (ISL) has an
extensive experimental railgun research program. The {\sc Pegasus} railgun 
installation consists out of a 10\,MJ capacitor based power supply and a 6\,m long
distributed energy supply (DES) railgun barrel. One focus of research
with this installation is the 
application of different armature concepts to improve the
armature/rail contact behavior \cite{peg_1}. A possible, military
application for powerful railguns is as artillery system. The availability
of a sufficient amount of electrical power on modern ships makes the
installation of electric guns on these ships an attractive possibility \cite{ship}. 
A military, fieldable gun system has to be as simple and robust as possible.
For a DES barrel, as currently being used in the {\sc Pegasus}
installation, current has to be routed to current injection points being distributed 
along the barrel. This makes these barrels mechanically complex and
adds more weight to the front part, increasing the time needed for pointing
movements. On the other hand, these barrels are more efficient than
comparable simple railgun barrels. In this paper a
simple railgun is compared to the DES railgun in order to be able to
asses how much better the DES railgun performs. For this a setup
closely resembling the current {\sc Pegasus} barrel was compared to a
simple railgun of the same size and connected to the same power
supply. The outcome of this study is used as a guide when deciding
on the architecture of a new barrel for the {\sc Pegasus} installation.

\section{Simple and DES Railgun}
\begin{figure}[b!]
\centering
\includegraphics[width=3.5in]{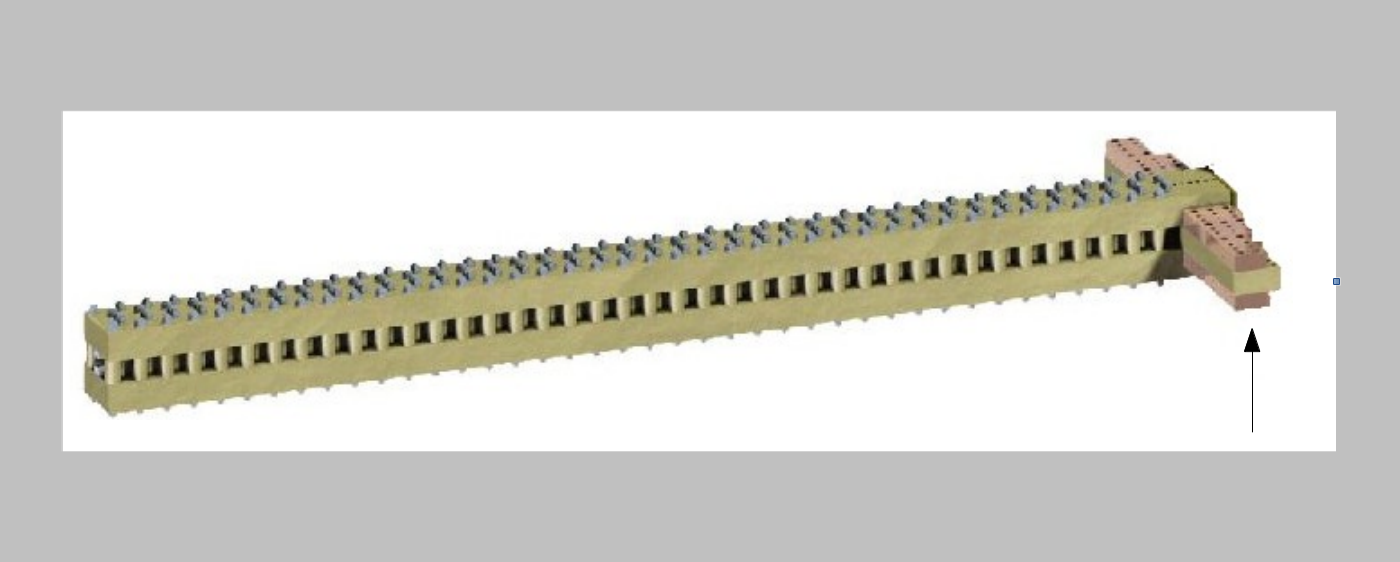}
\includegraphics[width=3.5in]{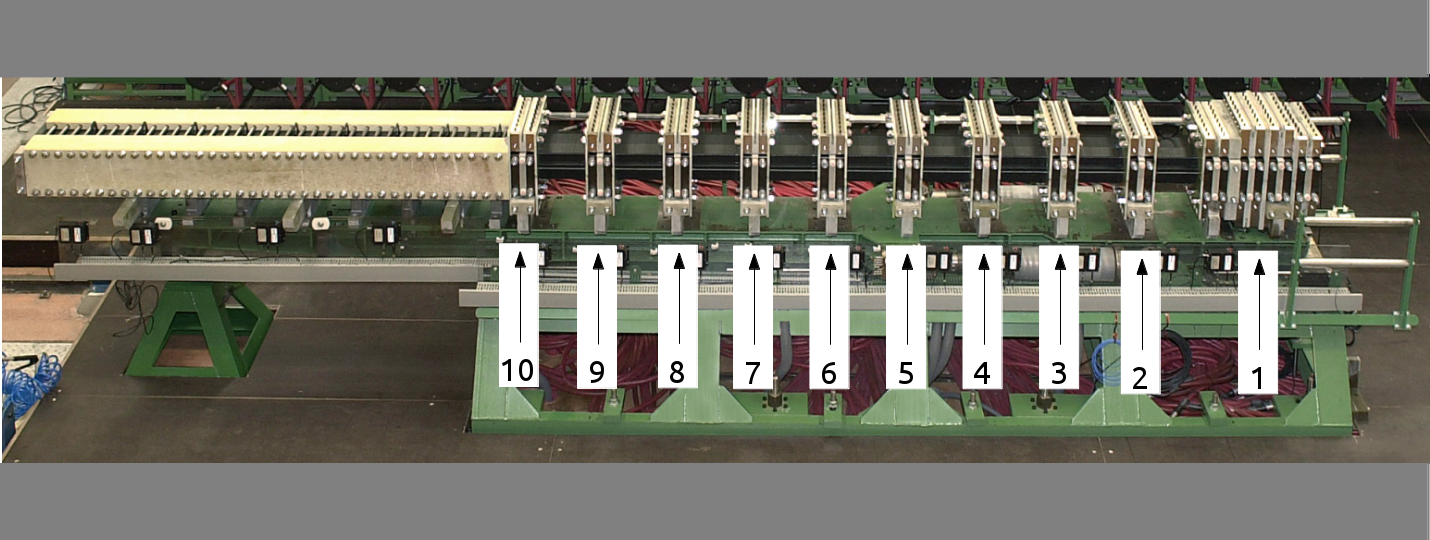}
\caption{Top: Simple railgun, the current injection bar is marked by
an arrow. Bottom: DES railgun with 10 current injection points.}
\label{fig_1}
\end{figure}

Electromagnetic railguns do exist in a wide variety of different
technical implementations \cite{green}. 
The simple or breech fed railgun is a straight forward
implementation of the railgun principle. It consists out of a housing, two
massive electrical conductors (the rails) and a current injection bar being
connected to the rails. A drawing of such a machine is shown in 
figure \ref{fig_1}(top). The current injection is located at the breech and
the armature is propelled by the electrical force towards the muzzle.
Seeing the simple railgun as an electrical circuit, the rails can be
described by a variable resistance and a variable inductance. The
actual value of the inductance $L$ and the resistance $R$ is calculated
by:
\begin{equation}
L=L' x \mbox{\,\,and\,\,} R=R' x
\label{eqn_1}
\end{equation}
with $x$ being the position of the armature and $L'$ and $R'$ are the
inductance and resistance gradients, respectively. The
equation \ref{eqn_1} show, that the resistance and the inductance of
the railgun grows with the progress of the armature through the
barrel. The energy lost resistively is not available for accelerating
the armature. The same is true for the magnetic energy being stored in
the inductance of the rails after the projectile has left the
launcher. Without further circuitry, this energy is dissipated
into heat by the muzzle flash. But, as demonstrated experimentally in 
\cite{4_stage_xram}, it is possible to recover the magnetic energy and
make it available for the next shot, thus allowing to increase the overall
energy efficiency. To reduce the resistive losses from the current
running through the length of the rails to the armature, the
distributed energy supply (DES) railgun places energy injection points
along the barrel. A \mbox{10-stage} DES railgun is shown in figure
\ref{fig_1}(bottom). The breech is to the right, the muzzle to the
left. The injections 1 to 10 are marked by arrows. In this figure the cables
connecting the energy storage with the railgun are not visible, as
they are not mounted. In a system with n stages, the stored electrical
energy is split into n units which can be triggered individually. When
the armature passes an injection point, it triggers the connected
energy supply unit. This reduces the path length the current 
has to travel from the injection point to the armature.  

\section{Simulations with {\sc Ngspice}}
\begin{figure}[tb!]
\centering
\includegraphics[width=3.5in]{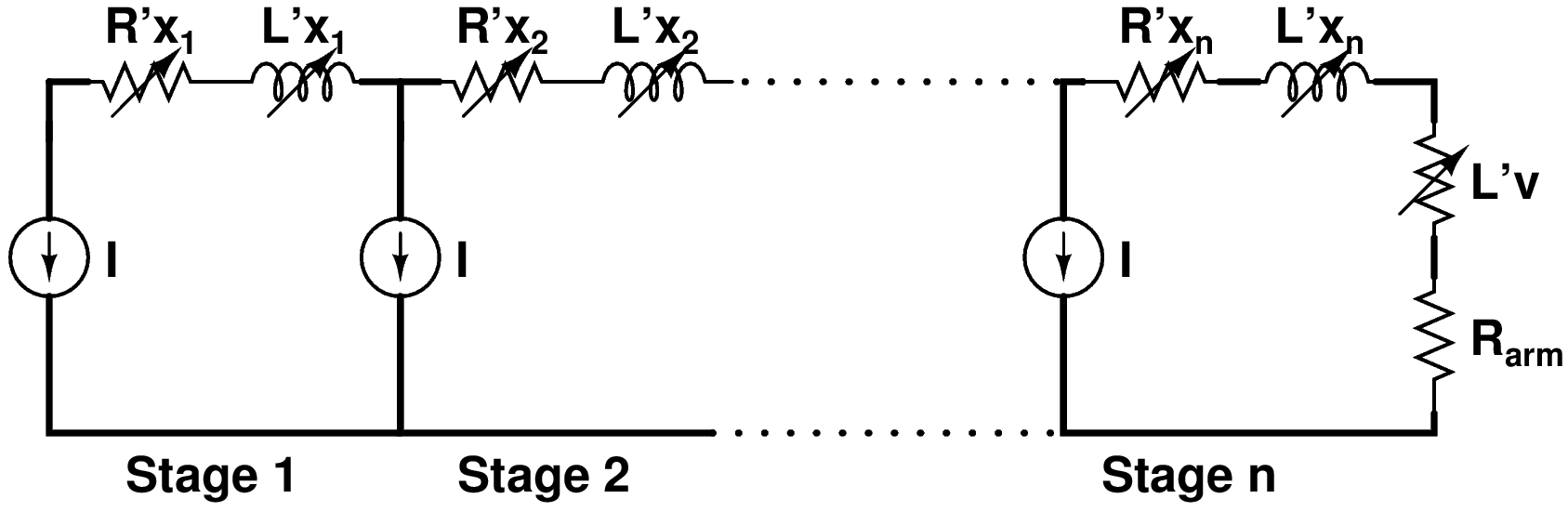}
\caption{Electrical circuit as implemented in {\sc Ngspice} for
simulating the railgun.}
\label{circuit_cc}
\end{figure}

Using {\sc Ngspice} (version 23) \cite{ngspice} the electrical circuit
for both railgun types were simulated. 
The
breech voltage of a railgun can be written as \cite{green} 
\begin{equation}
u_{breech}=L'x\frac{dI}{dt} + IL'v + IR'x + u_{muzzle}
\label{eqn_2}
\end{equation}
The first term on the right side describes the back electromotive force
of the inductance of the railgun. The second term is the so called "speed
voltage", the third is the voltage drop over the rail resistance and
the fourth the voltage drop over the armature including the contact
resistances. The individual components can be interpreted as
resistances by dividing by the current $I$. From this operation, the
electrical circuit shown in figure \ref{circuit_cc} is implemented into
{\sc Ngspice}. The back electromotive force is implicitly simulated
and is not represented by a circuit element. The element $R_{arm}$
includes the resistance from the armature and the contact resistance
from the armature to the rails. The dynamical variables of the
armature are derived by following the recipe as outlined in
\cite{yellow}. The acceleration is computed from the current using the railgun force law
\mbox{$F=1/2 L' I^2$}. The velocity and position of the armature are
calculated by integration. To make the simulation more realistic, mechanical 
friction was introduced by reducing the force acting on the projectile by 10\%. 
The circuit shown in figure \ref{circuit_cc} represents the DES
railgun. The values of the $x_n$ variables run from 0 to the length of
the stage n. As long as the armature has not yet entered a stage,
the value is 0 and when the armature leaves a stage the value stays at
the stage length. The more stages a DES setup has, the more the 
rail resistance seen by the current and therefore the energy lost by
ohmic heating of the rails is reduced. Another important consequence follows
from the electrical connection of the stages through the continuous rail. 
The magnetic energy stored in a stage the armature has left can discharge
in the remaining part of the railgun. This
reduces the amount of magnetic energy being stored in the railgun when the 
armature leaves the barrel and increases the efficiency. The electrical circuit 
of the simple railgun is gained by
reducing the circuit in figure \ref{circuit_cc} to the rightmost stage
("Stage n").

\subsection{Simulating the Pulsed Power Supply}
\begin{figure}[tb!]
\centering
\includegraphics[width=3.5in]{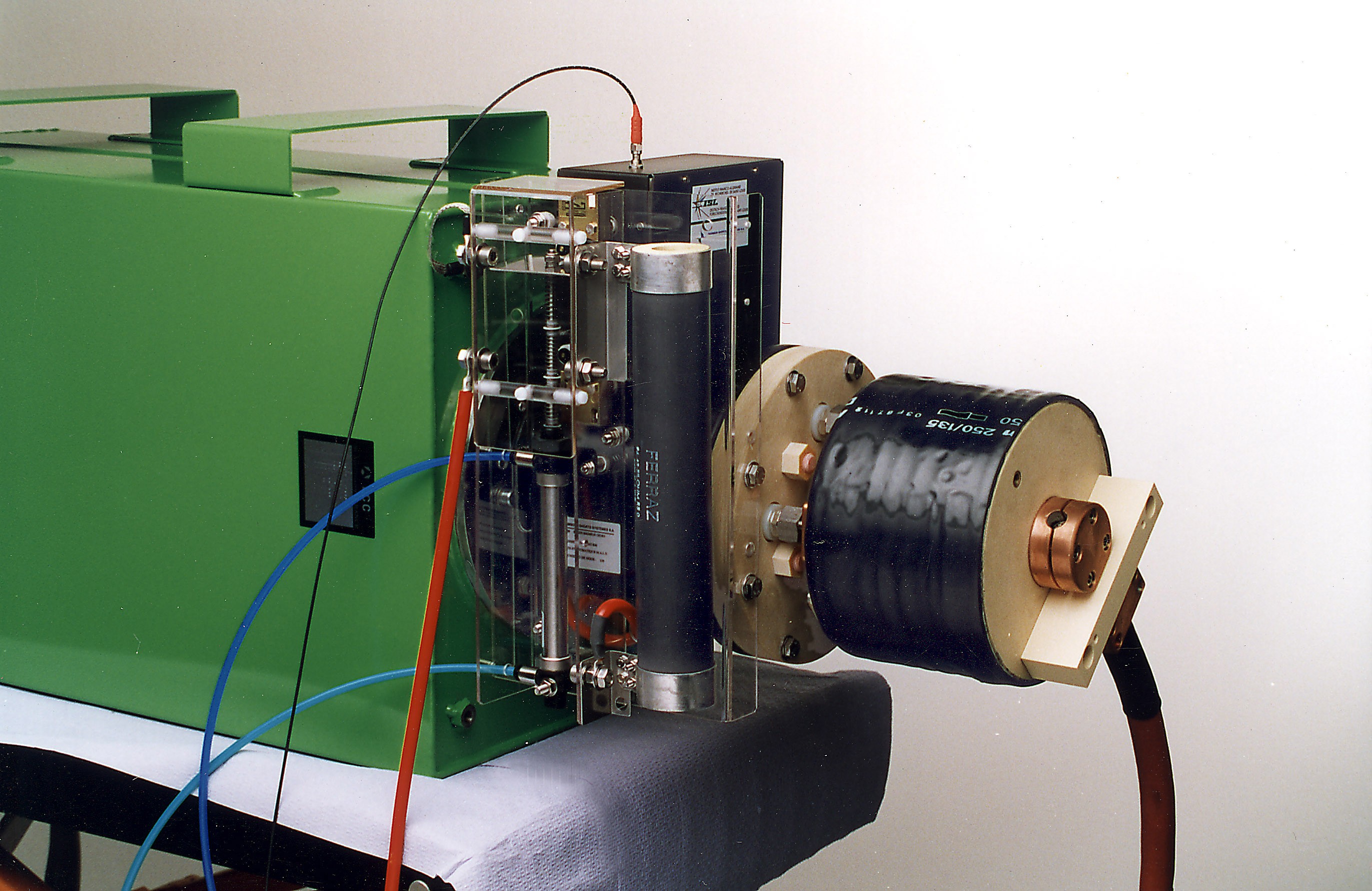}
\includegraphics[width=3in]{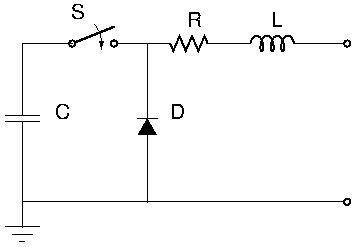}
\caption{Top: Capacitor module with 50\,kJ storage capacity. Bottom:
Electrical circuit for 50\,kJ module.}
\label{50kj}
\end{figure}
\begin{figure}[tb!]
\centering
\includegraphics[width=3.5in]{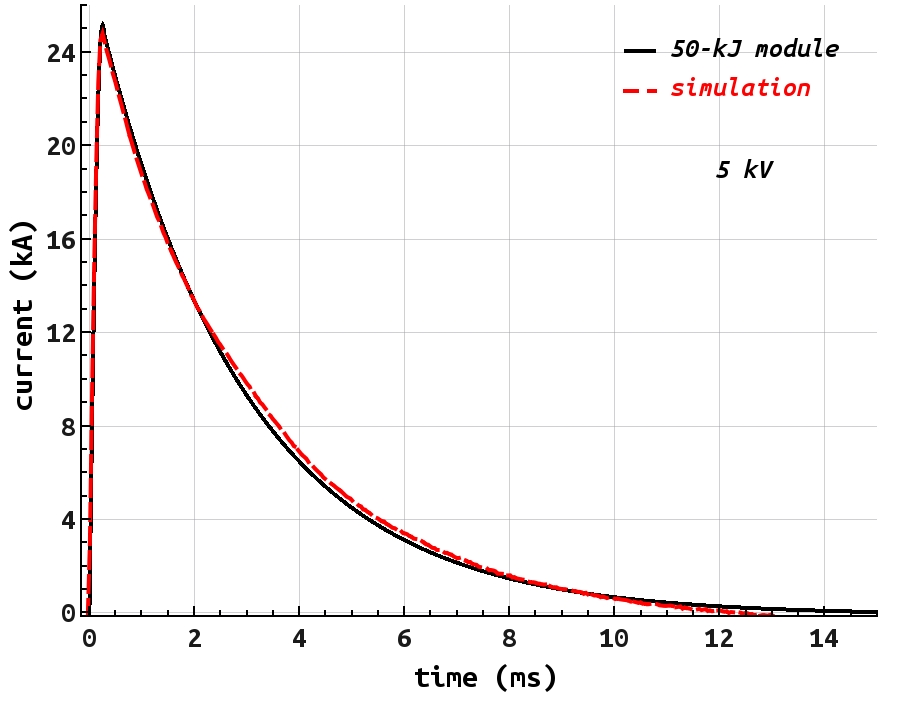}
\includegraphics[width=3.5in]{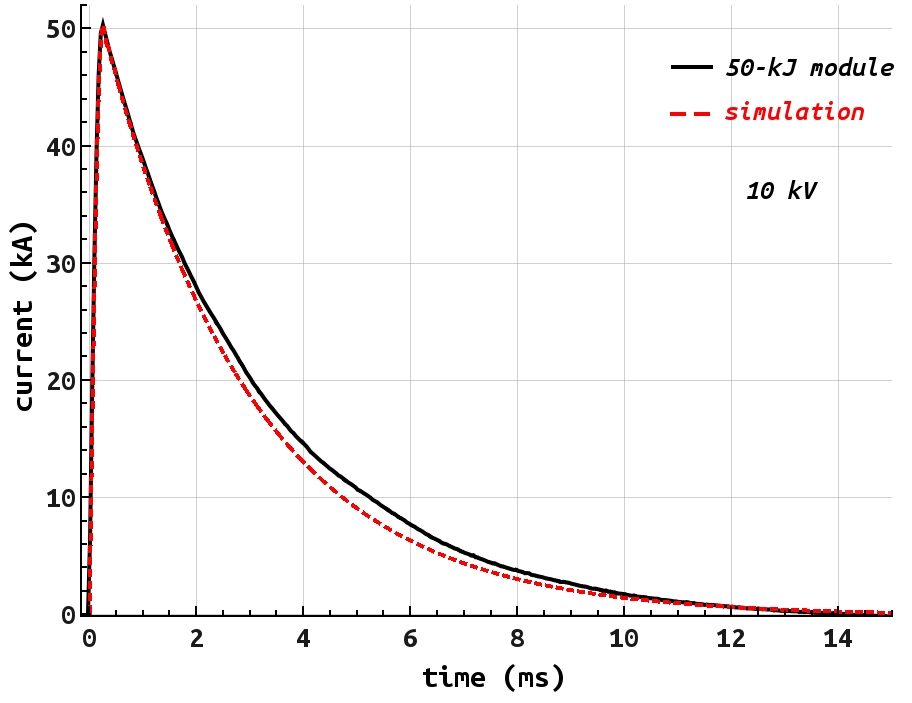}
\caption{Comparison of experimental short-circuit data and simulation
for a 50\,kJ capacitor module at a charging voltage of 5\,kV (top)
and 10\,kV (bottom).}
\label{short}
\end{figure}

For comparison of the two setups a 10\,MJ capacitor bank 
consisting out of 200 capacitor modules as shown in figure
\ref{50kj}(top) were simulated. Each module consists out of a 50\,kJ-capacitor,
a thyristor switch, a crowbar diode and a pulse forming coil. A detailed
description of the module and its performance is found in \cite{spahn_2}.
The figure \ref{50kj}(bottom) shows the electrical diagram as implemented
into {\sc Ngspice} to simulate the module. To allow for better
convergence of the Spice algorithm, the electrical circuit was simplified.
Instead of implementing the individual resistances and inductances of
all the elements, only one resistor R and one coil L was used. 
This is not fully describing the experimental situation, as the
switching to the crowbar circuit path reduces the resistance by
approx. 0.6\,m$\Omega$ and the inductance by 0.2\,$\mu$H. But
simulations comparing a fully implemented 50\,kJ module and the
simplified circuit from above showed only minor differences in the
resulting pulse height and shape. To successfully describe the experimental
short-circuit data over the range of the capacitor charging voltage
from 4\,kV to 10\,kV, the following 
values for the circuit were chosen: C=0.86\,mF,
R=11m$\Omega$, L=30.8\,$\mu$H. As shown in figure \ref{short} the simulation does 
describe the experimental data in the charging voltage range from
5\,kV to 10\,kV quite well.

\subsection{Simulation and Results}
\begin{figure}[tb!]
\centering
\includegraphics[width=3.5in]{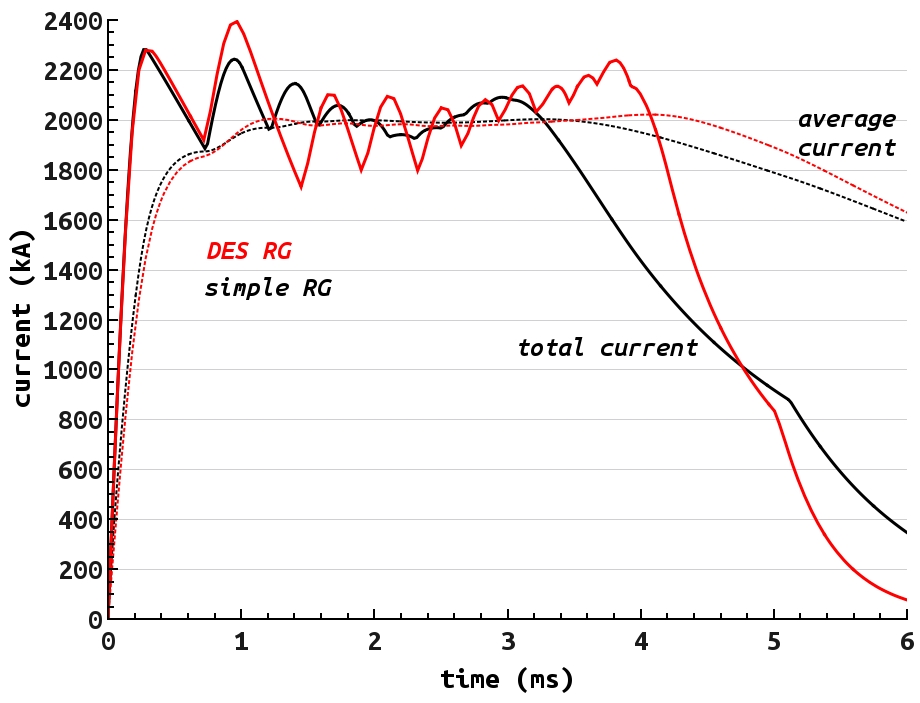}
\caption{Total current (solid lines) and average current (dotted line) 
distributions for the DES and simple railgun setup.}
\label{current}
\end{figure}
\begin{figure}[tb!]
\centering
\includegraphics[width=3.5in]{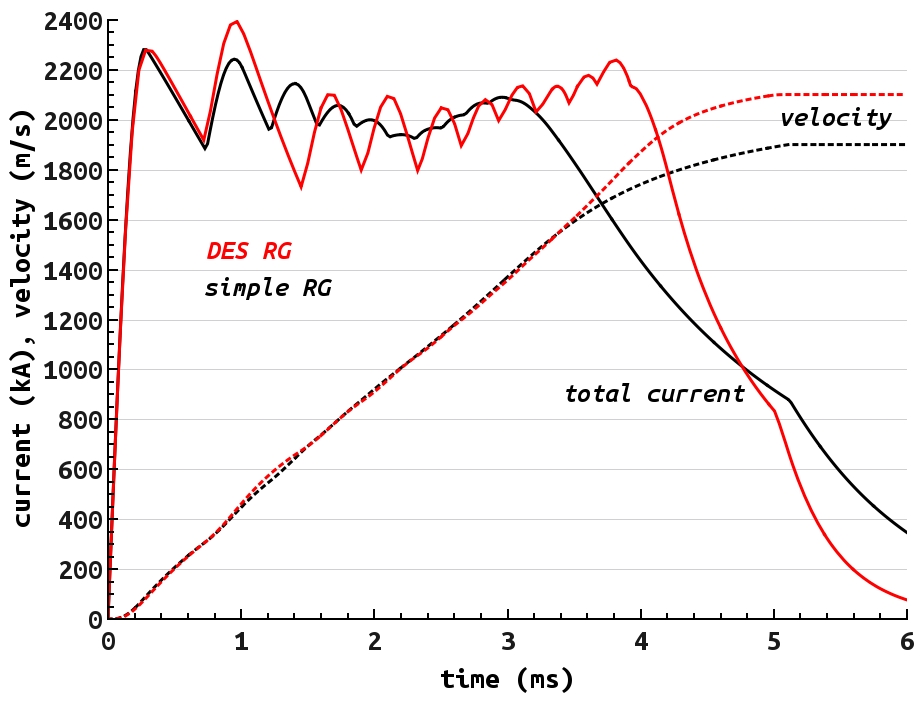}
\caption{Current and velocity distributions for DES and simple
railgun setups.}
\label{velocity}
\end{figure}
\begin{figure}[tb!]
\centering
\includegraphics[width=3.5in]{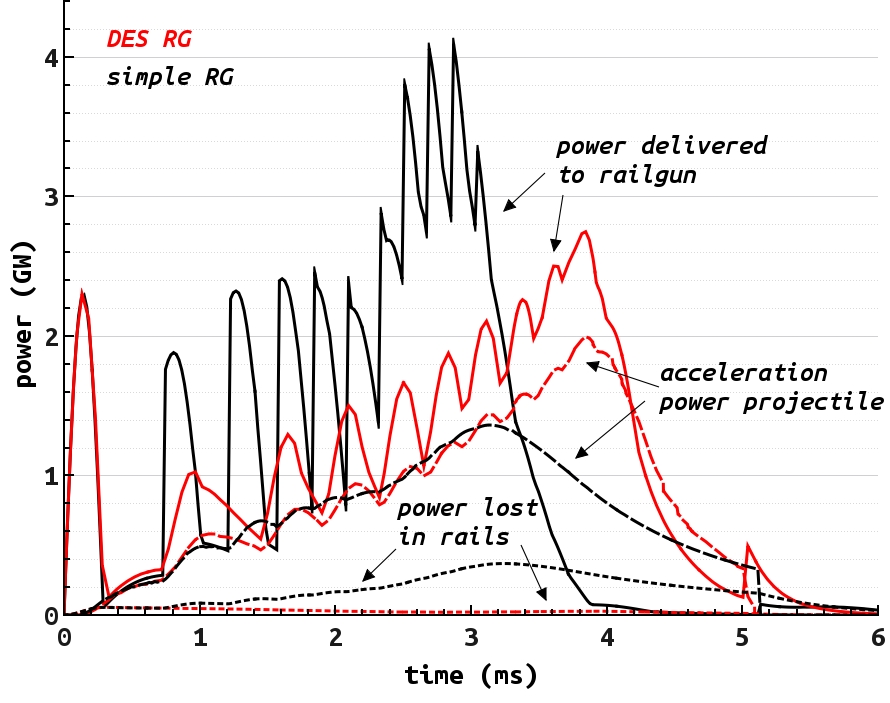}
\caption{Delivered electrical power to railgun, acceleration power for
the projectile and power lost in rails.}
\label{power}
\end{figure}
The energy being stored in the capacitors is dependent on the
charging voltage of the capacitors. For the simulations described
here, a voltage of 10.5\,kV, corresponding to an electrical energy 
of 9.45\,MJ was used. The 200 capacitor modules are arranged into 13
banks, each consisting out of 16 modules (with exception of the last
bank, which has only 8 modules). Each of the bank can be triggered
individually. The end-velocity of the projectile is
proportional to the action integral. Therefore one strategy to
optimize the velocity potential of a railgun is to arrange the
discharge sequence of the capacitor banks with respect to maximizing
the action integral. But, there are two important constraints for a
real railgun. The housing material has to withstand the repulsion
forces from the rails and the rail current carrying capacity is limited. Therefore, 
to obtain optimal performance a gun has to be driven close to its mechanical and 
current amplitude limit. A flat current profile at maximum current makes optimal use of the
available acceleration length. In a DES system, an approximate flat
current profile is achieved by using equally spatial spaced current
injections. As the instantaneous current distribution shows peaks whenever a bank is
triggered, the average current was calculated using
\begin{equation}
I_{av}=\frac{1}{t} \int_0^t I dt
\end{equation}
The distance in between the current injection points was varied to obtain a
flat shape of the $I_{av}$ trace. As a result for the DES setup, the 10 
current injection points were
distributed with equal distance to each other over the first 3.8\,m of the acceleration length.
An average current level of 2\,MA was adjusted by selecting
the capacitor charging voltage to 10.5\,kV. For the simple, breech fed
railgun setup the trigger instances of the capacitor banks
were adjusted for the same level of average
current for as long as possible during the acceleration. A good fit
could be achieved by triggering the banks when the projectile passes
equally spaced positions along the first 2.8\,m of the 6\,m long barrel.
Figure \ref{current} shows the current and average current distributions for
the simple and the DES railgun setups. The average current
distributions are quite the same up until 3.3\,ms, when the simple
railgun average current falls off, while for the DES railgun the
current plateau level is kept until 4.2\,ms.. Inspecting the total current
traces, both show initially (until approx. 1.2\,ms) the same amplitude
and behavior. Afterwards, one can see the effect of the growing
inductance and resistance of the simple railgun versus the DES setup.
The L-R combination acts as a variable low-pass filter with a lower
cut-off frequency for the simple railgun. As a result the current
trace becomes smoother with progressing acceleration time. The, on
average larger resistance in the simple setup results in higher ohmic
losses in the rails, but more importantly, the larger inductance
translates in more energy being stored in the magnetic field generated
by the rails. Both effects lead to an earlier depletion of the energy
supply for the simple railgun. 
Shot-out of the projectile is
at 5\,ms for the DES and at 5.2\,ms for the simple railgun setup. The
projectile velocity is shown in figure \ref{velocity}. As
the average current is the same up to 3.6\,ms, the velocity 
is the same for both setups until this time, as well.
Only afterwards, the DES system is performing better.
While the DES railgun projectile reaches 2100\,m/s, the simple railgun
accelerates the projectile to 1900\,m/s. It is interesting to note
that in the case of the simple railgun a significant part of the
acceleration is driven by the magnetic field stored in the railgun
itself. At 4.2\,ms the current starts to drop and the magnetic energy
stored in the rails of the simple railgun is at its maximum. At that 
point in time, the projectile has reached a velocity of 1500\,m/s and
has passed about half of the barrel length. During the remaining
barrel length the decaying magnetic field further accelerates the projectile 
to 1900\,m/s. 
\subsection{Delivered Power}
Even so the energy being stored in the capacitor banks is
the same for both setups, the required power to drive the projectile
through the different setups is very different, as shown in figure
\ref{power}. For the simple railgun, the release of the energy from
one bank leads to a strong peak and the width of the peaks narrows
with increasing velocity.  At maximum, the capacitor banks deliver 
close to 4.2\,GW. To sustain the current level of
approx. 2\,MA, the banks need to be fired in close proximity in time.
For the DES setup the maximum power for the banks stay below
2.8\,GW. Mainly two effects drive the voltage which the power supply
must overcome: The back electromotive force and the speed
voltage (the first and second term in equation \ref{eqn_2}).
The speed voltage is dependent on the velocity and the inductance
gradient. For both setups, the acceleration is nearly constant, therefore 
this voltage simply grows linearly with the acceleration time.
For the back electromotive force one needs to take into account the
different inductance seen by the banks, which are triggered in succession. 
For one bank with 16 capacitor modules in parallel, the inductance
including the cables is 1.9$\mu H$. For the DES setup the contribution
to the inductance due to the rails increases with the projectile
distance from the injection points. As the banks are triggered when an
injection point is passed by the projectile, this inductance 
contribution is limited to approximately the distance in between
the injection points (corresponding to about 0.2$\mu H$).
For the simple railgun, the additional inductance from
the rails increases up to the point when the last bank has fired at
half of the acceleration length.  This adds 1.5$\mu H$, thus nearly doubling the 
inductance seen by a capacitor bank. Therefore, in the simple
railgun case, the back electromotive force 
term grows to about twice its value at projectile start. This results in an
earlier depletion of the energy supply for the simple railgun. In addition 
to the power delivered from the capacitor banks
to the railgun, the power being converted into kinetic energy of the
2\,kg projectiles and the power lost across the rail resistance is shown
the same figure. In the case of the simple railgun, the acceleration power, 
reaches a maximum of 1.4\,GW. After the power
supply is depleted, the acceleration power is fed from the decay of the magnetic
energy being stored in the railgun. This effect compensates to a large
extend the drastic increase in power being delivered to the railgun
between 2.5\,ms and 3.4\,ms. For the DES railgun the acceleration
power is a large fraction of the delivered power for the whole time
period. As it is the intention, when implementing a DES system, the 
power lost in the rail resistance is negligible for the DES setup when
compared to the acceleration power. For the simple railgun the power
lost in rail heating is about 400\,MW, or 30\% of the acceleration power.
This contribution is still  small, when compared to the total power delivered to the railgun.  
\subsection{Efficiency}
\begin{figure}[tb!]
\centering
\includegraphics[width=3.5in]{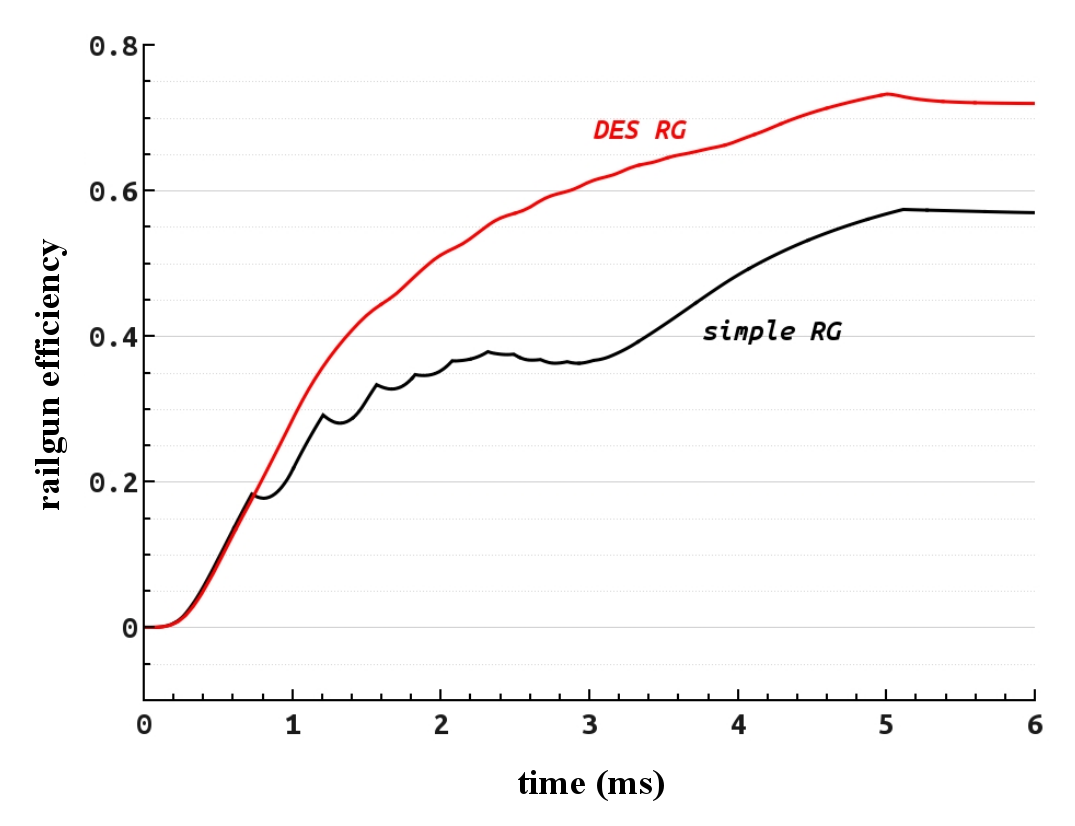}
\caption{Railgun efficiency for the DES and simple railgun setup.}
\label{railgun_eff}
\end{figure}
The efficiency of the railgun including the power supply is an
important design parameter. One can define the system efficiency as:
\begin{equation}
\eta = \frac{E_{kin}}{E_{cap}}
\label{eqn_3}
\end{equation}
Here $E_{kin}$ is the kinetic energy of the projectile and $E_{cap}$
is the energy initially stored in the capacitor banks. For the simple
railgun setup with a muzzle velocity of 1900\,m/s, the system
efficiency is $\eta = 38$\%, while the DES system reaches an efficiency $\eta = 46$\%.
It is common use to rate electrical generators or engines by a
efficiency that relates only to the machine, not taking into account
auxiliary systems. Therefore we define here a railgun efficiency
$\eta^{\star}$, that replaces the energy stored in the capacitor by
the energy delivered to the current injection points $E_{delivered}$ 
\begin{equation}
\eta^{\star}=\frac{E_{kin}}{E_{delivered}}
\label{eqn_4}
\end{equation}
For the two types of railguns figure \ref{railgun_eff} shows the
traces of the railgun efficiency. During the first phase of the
acceleration, the power supply keeps the current at the plateau level.
For the simple setup, this is until 3.3\,ms, for the DES setup until
4.2\,ms. Afterwards the current through the armature is supported by
the decaying magnetic field energy in the railgun rails. This
increases the efficiency, as it converts intermediately stored magnetic energy
into kinetic energy. The inductance of the simple railgun system
being used when the power supply units are exhausted is much larger
than for the DES setup. This effect increase the efficiency from 40\%
at the time when the power supply is depleted to 57\% at shot-out. For
the DES setup this increase is smaller, from 67\% to 73\%. Using the 
system efficiency and the railgun efficiency one can calculate the
efficiency for the power supply bank including the cables to 64\% for
the DES and 67\% for the simple railgun setup. The efficiency for the
simple railgun setup is higher, because the active time of the power supply
is shorter, slightly reducing the ohmic losses.

\section{Summary}
Using an existing 10\,MJ power supply two different railgun setups were
simulated using a {\sc Spice} algorithm. The performance of both were
investigated and compared. The main results of this study are
summarized in table \ref{tab_1}. For a given electrical energy the DES setup
is superior in all aspects concerning the electrical and dynamical
performance. It accelerates a 2\,kg mass to 2100\,m/s, this is 11\%
faster then the 1900\,m/s of the simple railgun. For the railgun
efficiency the difference is even larger, the DES setup reaches 73\%
which can be compared to 57\% for the simple railgun. But the details
of the acceleration process and the differences in power requirement
let this large difference shrink, when the overall system efficiency is
looked at: 38\% for the simple railgun and 46\% for the DES system.
This means that the efficiency difference is halved from a difference
of 16\% to "only" 8\%. This surprising effect comes from the fact,
that the power supply losses (including cables) are larger for the DES
setup than for the simple setup. As a conclusion, the simplicity of
the simple railgun setup, when connected to the existing power supply
does comes at a cost of velocity (-11\%) and efficiency (-8\%). These
differences are not large. Apart from these performance factors, there
are others, like cost of manufacture for the launcher and rails, 
serviceability, possibility to modify the launcher caliber or rail
dimensions and  accessibility of the acceleration volume. In the case
where a new lab-launcher needed to be designed, these
factors tipped the balance in favor of the simple railgun setup.

\begin{table}
\centering
\begin{tabular}[tbh]{|l|c|c|}
\hline
& Simple & DES \\
\hline
$E_{cap}$ & 9.45\,MJ& 9.45\,MJ \\
\hline
$E_{delivered}$ & 6.33\,MJ & 6.1\,MJ \\
\hline
velocity& 1900\,m/s & 2100\,m/s \\
\hline
$\eta$& 38\% & 46\% \\
\hline
$\eta^{\star}$& 57\% &73\% \\
\hline
\end{tabular}
\caption{Summary of the results for the simulations of a simple setup
and DES railgun setup.}
\label{tab_1}
\end{table}


\begin{thebibliography}{21}

\bibitem{peg_1}
S.\,Hundertmark, D.\,Simicic, G.\,Vincent,
\emph{Acceleration of Aluminum Booster Projectiles With PEGASUS},
IEEE Transactions on Plasma Science, Vol.\,43, No.\,5, May 2015.

\bibitem{ship}
S.\,Hundertmark, D.\,Lancelle,
\emph{A Scenario for a Future European Shipboard Railgun},
IEEE Transactions on Plasma Science, Vol.\,43, No.\,5, May 2015.

\bibitem{green}
W.\,Ying, R.\,A.\,Marshall, C.\,Shukang,
\emph{Pyhsics of Electric Launch}, Science Press, Beijing, 2004,
ISBN 7-03-012821-4.

\bibitem{4_stage_xram}
O.\,Liebfried and V.\,Brommer,
\emph{A Four-Stage XRAM Generator as Inductive Pulsed Power Supply for
a Small-Caliber Railgun},
IEEE Transactions on Plasma Science, Vol.\,41, No.\,10, October 2013.

\bibitem{ngspice}
NGSPICE is a Spice simulation program under the BSD license.
\it{http://ngspice.sourceforge.net/index.html}.

\bibitem{spahn}
E.\,Spahn, M.\,Lichtenberger, F.\,Hatterer,
\emph{Pulse forming network for the 10 MJ-railgun PEGASUS},
5$^{th}$ European Symposium on Electromagnetic Launch Technology,
Toulouse, France, 10--13 April 1995.

\bibitem{yellow}
R.\,A.\,Marshall, W.\,Ying,
\emph{Railguns: their Science and Technology},
China Machine Press, Beijing, 2004, ISBN 7-111-14013-3.

\bibitem{spahn_2}
E.\,Spahn, G.\,Buderer,
\emph{A flexible pulse power supply for EM- and ETC-launchers}, 
12$^{th}$ IEEE International Pulsed Power Conference, 1999. Digest of Technical
Papers, Vol.\,2, 27-30 June 1999.

\end{thebibliography}
\end{document}